\documentclass[review]{elsarticle}
\usepackage[section]{placeins}
\usepackage{lineno,hyperref}
\usepackage{graphicx}
\usepackage{booktabs}
\usepackage[dvipsnames]{xcolor}
\usepackage{tablefootnote}
\usepackage{multirow}
\usepackage{subfigure}
\usepackage{amsfonts}
\usepackage{soul}
\usepackage{array}
\usepackage{colortbl}
\usepackage{amsmath, amsthm, amssymb}
\usepackage{lipsum}
\usepackage{comment}
\usepackage[linesnumbered,ruled,vlined]{algorithm2e}
\usepackage{bm}
\usepackage{enumitem}
\theoremstyle{definition} 
\newtheorem{problem}{Definition}

\usepackage[noend]{algpseudocode}

\journal{Journal}

\begin{document}

\begin{frontmatter}

\title{Hierarchical Attentive Knowledge Graph Embedding for Personalized Recommendation
}


\author[mymainaddress]{Xiao Sha}
\ead{SHAX0001@e.ntu.edu.sg}
\author[mysecondaryaddress]{Zhu Sun\corref{mycorrespondingauthor}}
\ead{z.sun@mq.edu.au}
\author[mymainaddress]{Jie Zhang}
\ead{ZhangJ@ntu.edu.sg}

\address[mymainaddress]{School of Computer Science and Engineering, Nanyang Technological University, Singapore}
\address[mysecondaryaddress]{Department of Computing, Macquarie University, Australia}


\cortext[mycorrespondingauthor]{Corresponding author}

\begin{abstract}
Knowledge graphs (KGs) have proven to be effective for high-quality recommendation, where the connectivities between users and items provide rich and complementary information to user-item interactions. Most existing methods, however, are insufficient to exploit the KGs for capturing user preferences, as they either represent the user-item connectivities via paths with limited expressiveness or implicitly model them by propagating information over the entire KG with inevitable noise. \textcolor{black}{In this paper, we design a novel hierarchical attentive knowledge graph embedding (HAKG) framework to exploit the KGs for effective recommendation. Specifically, HAKG first extracts the expressive subgraphs that link user-item pairs to characterize their connectivities, which accommodate both the semantics and topology of KGs. The subgraphs are then encoded via a hierarchical attentive subgraph encoding to generate effective subgraph embeddings for enhanced user preference prediction.} Extensive experiments show the superiority of HAKG against state-of-the-art recommendation methods, as well as its potential in alleviating the data sparsity issue.
\end{abstract}



\end{frontmatter}


\section{Introduction}

Recommender systems have been widely applied to ease the information explosion problem by providing personalized item recommendation~\cite{sun2019research}.  Much effort has been devoted to the traditional collaborative filtering (CF)~\cite{NGCF19}, which however suffers from the sparsity of user-item interactions and the cold start problem~\cite{wang2018ripplenet}. To alleviate these issues, knowledge graphs (KGs)~\cite{sun2018recurrent} have been incorporated into recommender systems as 
the auxiliary data sources,
such as item attributes and user profiles~\cite{wang2019kgat}.  By exploring the interlinks of KGs, the connectivities between users and items help reveal their underlying relationship, which are complementary to the user-item interactions~\cite{wang2018explainable}. Till now, two types of KG-aware recommendation algorithms have been broadly studied, namely path-based~\cite{wang2018explainable} and propagation-based methods~\cite{wang2019kgat,wang2018ripplenet}. Despite of their success, the former methods represent the user-item connectivities via linear paths with limited expressiveness, which are insufficient to capture the rich semantics and the topology of KGs. The latter methods model such connectivities by propagating information over the entire KG, which inevitably introduce noise that is irrelevant to the specific user-item connectivity, thus misleading the inference of user preferences.  

\begin{figure}[t]
\centering
\includegraphics[width =0.6\textwidth]{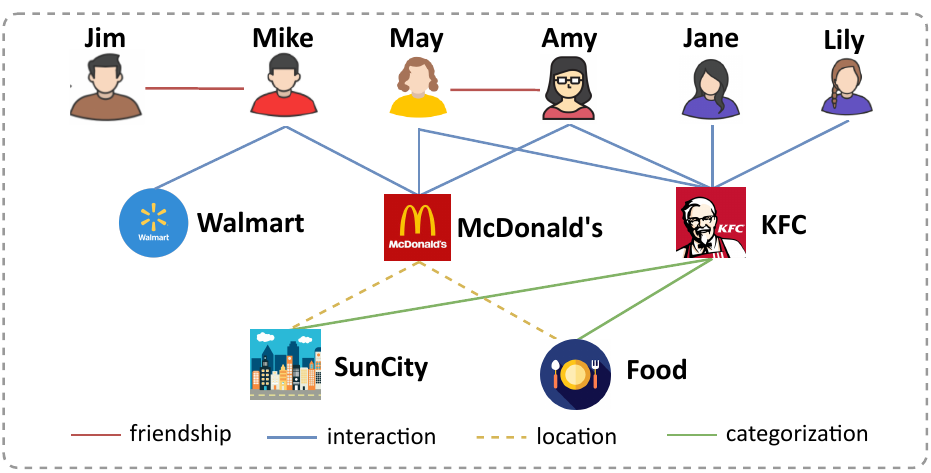}
\vspace{-0.05in}
\caption{A toy example of the KGs in Yelp, which contains users, business, categories, cities as entities; interaction, friendship, location, categorization as entity relations.
}\label{fig:toy_example}
\end{figure}

\textbf{Toy Example. }Figure \ref{fig:toy_example} depicts a toy example of the KGs in business domain \textcolor{black}{(i.e., Yelp)}. To infer Mike's preference over KFC, path-based methods extract separate paths that link Mike and KFC with length constraint (e.g., $\leq 3$) to model their connectivity, such as `Mike $\rightarrow$ McDonald's $\rightarrow$ May $\rightarrow$ KFC'. Nevertheless, we argue that such linear paths can only capture the partial semantics of the user-item connectivity, which is originally expressed by the subgraph with rich semantics and non-linear topology. For instance, \textcolor{black}{the subgraph between Mike and KFC reveals the strong interest of Mike over KFC. This is evidenced by the composite relations between KFC and McDonald's that Mike has interacted with before: belong to the same category (i.e. Food), located in the same city (i.e. SunCity), and simultaneously rated by two friends (i.e. May and Amy).} Due to the limited expressiveness of paths in describing the user-item connectivities, path-based methods fail to exploit the semantics and topology of KGs for capturing user preferences. 
Instead of directly representing the user-item connectivities, propagation-based methods implicitly model them by propagating information over the entire KG, where the user preferences are generally learned by aggregating information from all the neighbors. For instance, the preference of Mike is estimated by the aggregation of information from all his neighbors (e.g., Walmart, McDonald's and Jim), which however, may introduce noise (e.g., Walmart and Jim) that is irrelevant to the specific connectivity between Mike and KFC, thus misleading the inference of Mike's tastes over KFC. Due to the inevitable noise from the entire KG, propagation-based methods are suboptimal to characterize the pair-wise user-item connectivities for accurate user preference prediction.


To ease these issues,
we propose a novel framework named \textit{\underline{h}ierarchical \underline{a}ttentive \underline{k}nowledge \underline{g}raph embedding} (HAKG) to exploit the KGs for enhanced recommendation. In particular, HAKG explores the subgraphs that connect the user-item pairs in KGs for characterizing their connectivities, which is conceptually advantageous to most existing methods in that: (1) as a non-linear combination of separate paths, the subgraph contains both the rich semantics and topology of KGs, which is more expressive than linear paths; and (2) the subgraph only preserves entities and relations that are relevant to the specific user-item connectivity, which is able to avoid introducing noise from the entire KG. 

With these advantages in mind, HAKG aims to effectively encode the comprehensive information of subgraphs into low-dimensional representations (i.e., embeddings) for better revealing user preferences. The subgraph encoding is achieved via a hierarchical attentive embedding learning procedure with two core steps: (1) entity embedding learning, which learns embeddings for entities in the subgraph with a layer-wise propagation mechanism. In particular, each layer updates an entity's embedding based on the semantics propagated from its neighbors, and multiple layers are stacked to encode the subgraph topology into the learned entity embeddings; and (2) subgraph embedding learning, which attentively aggregates the entity embeddings to derive a holistic subgraph embedding. We deploy a novel self-attention mechanism to discriminate the importance of entities in the subgraph, so as to learn an effective subgraph embedding for better representing the user-item connectivity.

To summarize, our main contributions lie in three folds:
\begin{itemize}[topsep=0pt,itemsep=0pt,parsep=0pt,partopsep=0pt, leftmargin=*]
\item We propose to leverage the expressive subgraphs for better characterizing the user-item connectivities, so as to remedy the shortcomings of both path-based and propagation-based recommendation methods with KGs.
\item We develop a novel framework HAKG, which effectively encodes the subgraphs between user-item pairs via a hierarchical attentive embedding learning procedure. \textcolor{black}{By doing so, both the semantics and topology of KGs are fully exploited for enhanced item recommendation. }

\item We conduct extensive experiments on three real-world datasets. The results demonstrate the effectiveness of HAKG over the state-of-the-art methods in terms of recommendation performance, especially for \textcolor{black}{inactive users with sparse interactions over items.}
\end{itemize}


\section{Related Work}
\textcolor{black}{
KGs have become an increasingly popular data source leveraged
in a broad spectrum of disciplines, such as human-level intelligence~\cite{susanto2020hourglass} and sentiment analysis~\cite{cambria2020senticnet}, with recommendation being not exception. Comprehensive overviews on representation, acquisition and applications of KGs can be found in the survey~\cite{ji2020survey}.} In this section, we mainly focus on reviewing the KG-aware recommender systems, which incorporate the KGs to enhance recommendation. They can be generally classified into three categories: direct-relation based, path-based and propagation-based methods.
%
\textcolor{black}{
In addition, a brief review on studies which connect recommender systems with other related topics, i.e., cognitive models, binary codes and sentiment analysis, is provided as well.
}

\smallskip\noindent\textbf{Direct-relation based Methods.}
A line of research captures the direct relations between connected entities in KGs for entity embedding learning. Most of these methods are built on the conventional translation-based embedding techniques. 
For instance, KTUP~\cite{cao2019unifying} jointly learns the recommendation
model and the KG completion task based on TransH~\cite{wang2014knowledge}. \textcolor{black}{ACAM~\cite{yang2020knowledge} incorporates the KG embedding task via TransH to learn better item attribute embeddings.} RCF~\cite{xin2019relational} models the user preferences and item relations via DistMult~\cite{yang2014embedding}. \textcolor{black}{Similarly, JNSKR~\cite{chen2020jointly} enhances DistMult with non-sampling strategy to achieve efficient model training.} 
\textcolor{black}{Many researchers also incorporate KGs into sequential recommenders via conventional embedding techniques. For example, KERL~\cite{wang2020kerl} is a reinforcement learning based sequential recommender, which encodes the KGs via TransE to enhance state representations. MKM-SR~\cite{meng2020incorporating} involves KG embedding learning as an auxiliary task to promote the major task of sequential recommendation. Chorus~\cite{wang2020make} leverages TransE to model the item relations in KGs and integrates with their temporal dynamics for item embedding learning.} 

Though significant improvements have been achieved, these methods generally fail to capture the complex semantics of user-item connectivities for enhanced embedding learning, as they only consider the direct relations between entities in the KGs.

\smallskip\noindent\textbf{Path-based Methods.} Many methods explore the linear paths that connect entity pairs in the KGs to improve recommendation performance. For instance, PER~\cite{yu2014personalized}, and SimMF \cite{shi2016integrating} represent the user-item connectivities by defining different meta-path patterns. 
HERec~\cite{shi2019heterogeneous} and HINE~\cite{huang2017heterogeneous} leverage a meta-path based random walk strategy to generate path instances as the input of embedding learning process.  \textcolor{black}{HAN~\cite{wang2019heterogeneous} aggregates features from meta-path based neighbors to learn entity embeddings
via node-level and semantic-level attentions. However, this method discards
all intermediate entities along the meta-path by only considering two
end entities, which results in information loss.} \textcolor{black}{HCDIR~\cite{bi2020heterogeneous} leverages the meta-path based neighbors to enhance user embedding learning for cross domain
insurance recommendation. ACKRec~\cite{gong2020attentional} generates entity embeddings based on meta-paths to assist knowledge concept recommendation in MOOCs platform. MetaHIN~\cite{lu2020meta} incorporates meta-paths into a meta-learning framework for cold-start recommendation. {NIRec}~\cite{jin2020efficient} learns interactive patterns between meta-paths with fast Fourier transform to improve CTR prediction.} However, defining effective meta-paths requires domain knowledge, which can be rather labor-intensive for complicated KGs with diverse entities and relations. 

To address this limitation, the recently proposed RKGE~\cite{sun2018recurrent} and KPRN~\cite{wang2018explainable} automatically extract the paths linking user-item pairs with length constraint, and then model these paths via recurrent neural networks (RNNs). Later, KARN~\cite{zhuknowledge2020} encodes the paths between users and items with RNNs and attention networks to assist in click-through rate (CTR) prediction. This method, however, relies on the additional information from users’ clicked history sequences and textual information of items. \textcolor{black}{Recently, ADAC~\cite{zhao2020leveraging} supervises the path finding in KGs via reinforcement learning algorithms to help achieve fast convergence and improve explainability. KGPolicy~\cite{wang2020reinforced} explores the paths in KGs by employing a reinforcement learning agent for high-quality negative sampling. Besides, Fu et al.~\cite{fu2020fairness} proposes a fairness-aware path reranking algorithm for explainable recommendation.}

Despite of the effectiveness, path-based methods generally suffer from the limited expressiveness of linear paths in describing the complicated user-item connectivities. In contrast, our HAKG explores the expressive subgraphs that link the user-item pairs to represent their connectivities, so as to exploit the semantics and topology of KGs for revealing user preferences.

\smallskip\noindent\textbf{Propagation-based Methods. }Recent methods model the user-item connectivities by propagating information over the entire KG. For instance, KGCN~\cite{wang2019knowledge} propagates the item information within KGs to generate better item embeddings via graph convolutional networks (GCNs)~\cite{hamilton2017inductive}. Most recently, KGAT~\cite{wang2019kgat} recursively propagates 
information from an entity's neighbors to refine its embedding, and an attention mechanism is utilized to discriminate the importance of neighbors. \textcolor{black}{KGRL~\cite{chen2020knowledge} encodes the KGs via GCNs to guide the reinforcement learning process for interactive recommendation.  KGSF~\cite{zhou2020improving} captures the semantic relations between words and items in KGs for conversational recommendation. KGQR~\cite{zhou2020interactive} propagates user preferences over KGs to solve the sample efficiency problem in interactive recommenders. }

Despite of the better exploration of the KG topology, the information diffusion over the entire KG would inevitably introduce noise that is irrelevant to the specific user-item connectivity, and thus adversely degenerates the embedding learning ability. \textcolor{black}{Moreover, instead of explicitly encoding the user-item connectivities, most these methods model them with merely user and item embeddings, and thus fail to fully exploit the connectivities for revealing user preferences.} In contrast, HAKG represent the user-item connectivities by constructing subgraphs with relevant entities and relations, which are further encoded by a hierarchical attentive embedding learning procedure for accurate user preference prediction. 

\noindent\textbf{Connecting Recommendation with Other Related Topics.} \textcolor{black}{Recommender systems could also be designed and evaluated considering cognitive factors. Angulo et al.~\cite{angulo2020bridging} proposed a special issue on bridging cognitive models and recommender systems, which highlights the advantages of bridging different fields for the study of cognitive architectures for recommendation, as well as recommender systems shaping cognitive architectures. Expert recommender systems can automate procedures and tasks in a cognitive manner by leveraging the assessment from collaborative human expertise. For instance, 
Yang et al.~\cite{yang2020doctor} leverage intuitionistic fuzzy sets (IFSs) for improving standard recommendations.  }
%
\textcolor{black}{In the fast-growing e-commerce scenario, hashing for recommendation has attracted increasing attention as binary codes can significantly reduce the storage and make calculations efficient. For example, Suthee et al.~\cite{li2019disentangled} build a variational deep semantic hashing (VDSH) model based on variational auto-encoders. Li et al.~\cite{li2019learning} propose a novel deep collaborative hashing (DCH) model that learns efficient binary codes for dealing with out-of-sample cases.}
\textcolor{black}{In addition, affective computing and sentiment analysis shows the great potential in intelligent systems. For instance, Bi et al.~\cite{bi2020crowd} conduct asymmetric impact-performance analysis for understanding customer satisfaction. Schuurmans et al.~\cite{schuurmans2020intent} leverage the hierarchical structure to help better understand customer intentions.  By revealing which features customers would like to enjoy, affective computing and sentiment analysis would play a key role in enhancing recommendations.}


\section{Hierarchical Attentive Knowledge Graph Embedding}

This section first provides the formulation of the investigated
task in this paper, and presents an overview of the proposed framework -- Hierarchical Attentive Knowledge Graph Embedding for effective recommendation. We then go into the details of the components of HAKG, and finally discuss the model optimization and model complexity. 

\begin{table}[t]
    \centering
    \renewcommand{\arraystretch}{1.2}
    \footnotesize
    \caption{Notations}
    \label{tab:notations}
    \begin{tabular}{>{\color{black}}l!{\color{black}\vrule}>{\color{black}}l}
    \arrayrulecolor{black}\specialrule{.15em}{.05em}{.05em}
    Notations &  Descriptions\\
    \specialrule{.05em}{.05em}{.05em}
    $\mathcal{U} = \{u_1, u_2, \cdots, u_{\vert\mathcal{U}\vert}\}$& User set\\\hline
    $\mathcal{I} = \{i_1,i_2, \cdots, i_{\vert\mathcal{I}\vert}\}$& Item set\\\hline
    $\mathbf{R}^{{\vert\mathcal{U}\vert} \times {\vert\mathcal{I}\vert}}$&User-item implicit feedback matrix\\\hline
    $r_{u,i},\widetilde{r}_{u,i}$&Observed and estimated ranking scores\\\hline
    $\mathcal{G}, \mathcal{G}_{(u,i)}$& Knowledge Graph and the subgraph for $(u,i)$\\\hline
    $\mathcal{E},\mathcal{E}_{(u,i)}$&Entity sets of $\mathcal{G},\mathcal{G}_{(u,i)}$\\\hline
    $\mathcal{L},\mathcal{L}_{(u,i)} $&Link sets of $\mathcal{G},\mathcal{G}_{(u,i)}$\\\hline
    $n=\vert \mathcal{E}_{(u,i)}\vert$&The number of entities in $\mathcal{G}_{(u,i)}$\\\hline
    $\mathcal{P}_{(u,i)}$ &The sampled path set for $(u,i)$\\\hline
    $K = \vert\mathcal{P}_{(u,i)}\vert$ & The number of sampled paths for $(u,i)$\\\hline
    $\mathbf{e}_h \in \mathbb{R}^{d_e},\mathbf{t}_h\in \mathbb{R}^{d_t}$ &Embeddings of entity $e_h$ and its type $t_h$ \\\hline 
    $\mathbf{r}_{h,k} \in \mathbb{R}^{d_r}$ &Embedding of the relation between $e_h$,  $e_k$\\\hline
    $d_e,d_t,d_r$ & Embedding sizes of entity, entity type and relation\\\hline
    $\mathcal{N}_h$& Neighbors of entity $e_h$ in $\mathcal{G}_{(u,i)}$\\\hline
    $L$ & The number of propagation layers\\\hline
    $\mathbf{H}_{(u,i)}\in \mathbb{R}^{n\times d_e}$ & Entity embedding matrix of $\mathcal{G}_{(u,i)}$\\\hline
   $\mathbf{A}_{(u,i)}\in \mathbb{R}^{m\times n}$ & Attention weight matrix for entities in $\mathcal{G}_{(u,i)}$ \\\hline
   $m$& The number of attention heads\\\hline
    $\mathbf{e}_{u},\mathbf{e}_{i},\mathbf{g}_{(u,i)} \in \mathbb{R}^{d_e}$ & Embeddings of $u,i$ and $\mathcal{G}_{(u,i)}$\\
    \specialrule{.15em}{.05em}{.05em}
    \end{tabular}
\end{table}
\subsection{Task Formulation and HAKG Framework} 
We denote the user set and item set as $\mathcal{U} = \{u_1, u_2, \cdots, u_{\vert\mathcal{U}\vert}\}$ and $\mathcal{I} = \{i_1,i_2, \cdots, i_{\vert\mathcal{I}\vert}\}$, with $\vert\mathcal{U}\vert$ and $\vert\mathcal{I}\vert$ as the number of users and items, respectively. As we focus on the personalized ranking task, we binarize the user-item rating matrix into implicit feedback matrix $\mathbf{R} \in \mathbb{R}^{\vert\mathcal{U}\vert \times \vert\mathcal{I}\vert}$ by following state of the art~\cite{he2017neural, wang2019kgat}, where the entry $r_{u,i} = 1$ if the user $u$ has rated
the item $i$, and 0 otherwise. \textcolor{black}{Table \ref{tab:notations} summarizes all the notations utilized in this paper.}
For generality, we use `entity' to refer to the objects
(e.g., user, business, category, and city) that can be mapped into a KG (denoted as $\mathcal{G}$). The definitions of KGs and the investigated task are given as below. 
\begin{problem}{\textbf{Knowledge Graph}.}
Let $\mathcal{E, L}$ denote the sets of entities and links, respectively.   
A KG is defined as an undirected graph $\mathcal{G} = (\mathcal{E}, \mathcal{L})$ with entity type and link type mapping functions $\phi : \mathcal{E} \rightarrow \mathcal{A}$ and $\varphi : \mathcal{L} \rightarrow \mathcal{R}$.
Each entity $e \in \mathcal{E}$ belongs to an entity type $\phi(e) \in \mathcal{A}$, and each link $l \in \mathcal{L}$ belongs to a link type $\varphi(l) \in \mathcal{R}$. The types of entities $\vert \mathcal{A}\vert > 1$ or the types of links $\vert \mathcal{R}\vert > 1$ in KGs.
\end{problem}
\begin{problem}{\textbf{KG-aware Top-$N$ Recommendation.}}
Given the KG $\mathcal{G}$, for each user $u$ $\in$ $\mathcal{U}$, our task is to generate a ranked list of top-$N$ items that will be of interest to user $u$.
\end{problem}

\begin{figure*}[t]
\centering
  \includegraphics[width=1.04\textwidth]{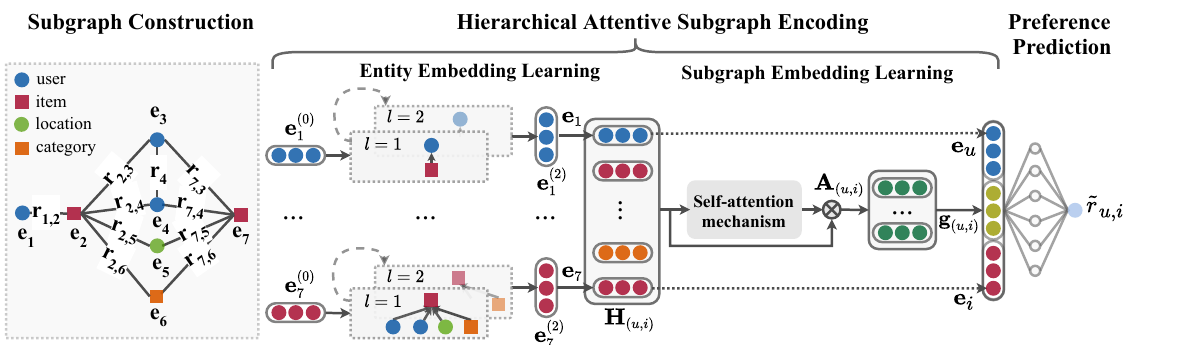}
  \vspace{-0.35in}
  \caption{The overall framework of HAKG describing the case of a user-item pair, which consists of three modules: Subgraph Construction, Hierarchical Attentive Subgraph Encoding and Preference Prediction. The Hierarchical Attentive Subgraph Encoding is composed of two core steps: Entity Embedding Learning and Subgraph Embedding Learning. 
  }\label{fig:framework}
  \vspace{-0.1in}
\end{figure*}

The overall framework of HAKG is illustrated by Figure \ref{fig:framework}, composed of three modules: (1) Subgraph Construction -- it automatically constructs the expressive subgraph that links the user-item pair to represent their connectivity; 
(2) Hierarchical Attentive Subgraph Encoding -- the subgraph is further encoded via a hierarchical attentive embedding learning procedure, which first learns embeddings for entities in the subgraph with a layer-wise propagation mechanism, and then attentively aggregates the entity embeddings to derive the holistic subgraph embedding; 
(3) Preference Prediction -- with the well-learned embeddings of the user-item pair and their subgraph connectivity, it uses non-linear layers to predict the user's preference towards the item.

\subsection{Subgraph Construction}
It is computationally prohibitive to construct subgraphs between user-item pairs via traditional graph mining methods, such as BFS ~\cite{zhou2006breadth} and DFS~\cite{tarjan1972depth} with an $O(\vert\mathcal{E}\vert + \vert\mathcal{L}\vert)$ complexity, where $\vert\mathcal{E}\vert$ and $\vert\mathcal{L}\vert$ denote the number of entities and links of the KG.
To this end, we propose a more efficient subgraph construction strategy to reduce the complexity, which converts the subgraph mining into path sampling and then reconstructs the subgraphs by assembling the sampled paths between user-item pairs.

\smallskip\noindent\textbf{Path Sampling. }It is not feasible to extract all the paths between user-item pairs from the KG, since the number of paths grows exponentially with the increasing of path length~\cite{wang2018explainable}. 
Actually, paths with a short length (no more than six) are sufficient to model the user-item connectivity, whereas long paths may bring in remote neighbors with noise~\cite{sun2018recurrent}. 
\textcolor{black}{For efficiency, we uniformly sample $K$ paths, each with length up to six, that connect a user-item pair. We take a similar approach as DeepWalk~\cite{perozzi2014deepwalk} for path sampling. Specifically, starting from the user $u$, we conduct random walks with a maximal depth of six, and only keep the paths that lead to the item $i$. As such, we generate a set of sampled paths for ${(u,i)}$ and denote the path set as $\mathcal{P}_{(u,i)}$, which can be done offline.} The impact of the number of sampled paths $K$ (i.e., $\vert \mathcal{P}_{(u,i)}\vert$) has been investigated in the experiments (\textit{cf.} Section \ref{parasensitivity}). Note that, we employ the uniform path sampling for simplicity and leave the exploration of non-uniform samplers (e.g., importance
sampling~\cite{wang2019knowledge}) as the future work.

\smallskip\noindent\textbf{Path Assembling. }By assembling the sampled paths in $\mathcal{P}_{(u,i)}$, we generate the subgraph $\mathcal{G}_{(u,i)}$ for $(u,i)$. In particular, we traverse $\mathcal{P}_{(u,i)}$ to map the objects and connections along a path into $\mathcal{G}_{(u,i)}$ as entities and links, respectively. Take the path: Mike $\rightarrow$ McDonald's $\rightarrow$ May $\rightarrow$ KFC in Figure \ref{fig:toy_example} as an example. The objects Mike, McDonald's, May, KFC are mapped as different types of entities: $e_1$, $e_2$, $e_3$, $e_7$; and the connections between entities (e.g., Mike and McDonald's) are mapped as links with relations (e.g., $r_{1,2}$). As we can see, $\mathcal{G}_{(u,i)}$ comprehensively integrates the semantics from entities, entity types and relations in KGs. which is then fed into the hierarchical attentive subgraph encoding module to learn a holistic subgraph embedding for characterizing the user-item connectivity.


%

\subsection{Hierarchical Attentive Subgraph Encoding}
Hierarchical attentive subgraph encoding learns effective subgraph embeddings for better representing the user-item connectivities. It however, is non-trivial due to (1) the rich semantics from the heterogeneous entities and relations of the subgraph, as well as the high-order subgraph topology; (2) the varying importance of entities in the subgraph for inferring the target user's preference. To address these challenges of subgraph encoding, we equip HAKG with a hierarchical attentive embedding learning procedure, which consists of two core steps: (1) entity embedding learning -- it exploits both the semantics and topology of the subgraph via a layer-wise propagation mechanism, so as to learn effective embeddings for entities in the subgraph; and (2) subgraph embedding learning -- it attentively aggregates the entity embeddings to derive the holistic subgraph embedding, where a self-attention mechanism is leveraged to discriminate the importance of entities.

\begin{figure*}[t]
\hspace{1.02in}
\includegraphics[width=0.9\textwidth]{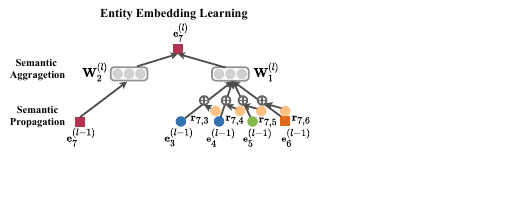}
\vspace{-1in}
\caption{\textcolor{black}{Illustration of the two major operations of entity embedding learning: semantics propagation and semantics aggregation.}
}\label{fig:entity_learning}
\end{figure*}

\subsubsection{\textbf{Entity Embedding Learning}}

\ 

\noindent Entity embedding learning generates effective entity embeddings by encoding both the semantics and topology of the subgraph. To achieve this, we leverage a layer-wise propagation mechanism, where each layer updates an entity's embedding based on the semantics propagated from its neighbors; and multiple layers are stacked to exploit the high-order subgraph topology. Specifically, there are three major operations: embedding initialization, semantics propagation and semantics aggregation, \textcolor{black}{as illustrated by Figure~\ref{fig:entity_learning}}.

\smallskip\noindent\textbf{Embedding Initialization. }We initialize the entity embeddings by incorporating the heterogeneous entity types.
Specifically, we first adopt an embedding look-up layer to project each entity $\mathbf{e}_h$ (e.g., KFC) and its corresponding entity type $t_h$ (e.g., business) with two low dimensional vectors
$\mathbf{e}_h \in \mathbb{R}^{d_e}$ and \textcolor{black}{$\mathbf{t}_h\in \mathbb{R}^{d_t}$, where $d_e$ and $d_t$ are the respective embedding sizes.} The two vectors are then fused as the initial embedding for $e_h$, to be refined by the layer-wise propagation process, denoted as:
\begin{equation}\label{equ:typeembedding}
\small
    \mathbf{e}^{(0)}_h = f (\mathbf{e}_h \oplus \mathbf{t}_h),
\end{equation}

\noindent where $\mathbf{e}^{(0)}_h$ is the initial embedding of $e_h$ at propagation layer $l=0$; $f(\mathbf{x})\!=\!\sigma \left(\mathbf{W}\mathbf{x} + \mathbf{b}\right)$, with $\mathbf{W}$ and $\mathbf{b}$ as the transformation matrix and bias term, respectively; and $\oplus$ denotes the concatenation operation. In this way, the heterogeneous type information is explicitly incorporated into entity embedding learning for better encoding the rich semantics within subgraphs.

\smallskip\noindent\textbf{Semantics Propagation. } 
The semantics
propagation operation models the semantics propagated from neighbors for the target entity. In KGs, the rich semantics from various relations (e.g. friendship, interaction, location) are critical for understanding the diverse user intents ~\cite{wang2018explainable}, which however are generally overlooked in most existing studies. To this end, we propose to explicitly model the heterogeneous relations during propagation for better encoding the semantics of subgraphs. In particular, we represent the relation between the target entity $e_h$ and its neighbor $e_k$ as $r_{h,k}$,\textcolor{black}{\footnote{For simplicity, we set $r_{h,k} = r_{k,h}$.}} and define the semantics to be propagated from $e_k$ to $e_h$ at layer $l$ as follows:
\begin{equation}
\small
\label{equ: propgation}
  \mathbf{s}^{(l)}_{h \leftarrow k} = \frac{1}{\sqrt{\vert \mathcal{N}_h\vert \vert \mathcal{N}_k}\vert} \left(\mathbf{W}^{(l)}_{1} \cdot (\mathbf{e}_k^{(l-1)} \oplus \mathbf{r}_{h,k}) \right),
\end{equation}

\noindent where $1/\sqrt{\vert \mathcal{N}_h\vert \vert \mathcal{N}_k}\vert $ is the graph Laplacian norm as used in GCNs, with $\vert \mathcal{N}_h\vert$ and $\vert\mathcal{N}_k\vert$ as the number of neighbors of $e_h$ and $e_k$ in the subgraph $\mathcal{G}_{(u,i)}$, respectively; $\mathbf{W}^{(l)}_{1}$ is the trainable weight matrix at propagation layer $l$; $\mathbf{e}_k^{(l-1)}$ is the embedding of neighbor $e_k$ generated from the $(l-1)$-th layer; $\mathbf{r}_{h,k} \in \mathbb{R}^{d_r}$ is the embedding of the linked relation $r_{h,k}$, with $d_r$ as the embedding size; and the concatenation of the two embeddings $\mathbf{e}_k^{(l-1)} \oplus \mathbf{r}_{h,k}$ encodes the semantics from neighbor $e_k$ along the relation $r_{h,k}$. Such composite semantics are further transformed into $\mathbf{s}^{(l)}_{h \leftarrow k}$, which will be propagated to the target entity $e_h$ for refining its embedding. In this way, HAKG sufficiently encodes the rich semantics of the subgraph from various entities and relations, contributing to a better entity embedding learning. 

\smallskip\noindent\textbf{Semantics Aggregation. }The semantics aggregation operation aggregates the semantics from neighbors to refine the target entity's embedding. Specifically, given $\mathbf{s}^{(l)}_{h \leftarrow k}$, we update the embedding of $e_h$ at propagation layer $l$ as follows: 
%

\begin{equation}\label{eq:semantic_agg}
\small
  \mathbf{e}_h^{(l)}= \sigma \left(\mathbf{W}^{(l)}_{2}\mathbf{e}_h^{(l-1)} + \sum\nolimits_{\forall e_k\in\mathcal{N}_h}\mathbf{s}^{(l)}_{h \leftarrow k}\right),
\end{equation}
\noindent where $\sigma$ is the activation function of ReLU; $\mathbf{W}^{(l)}_{2}\mathbf{e}_h^{(l-1)}$ preserves the information of $e_h$' embedding from previous layers, which is able to alleviate the over-smoothing issue by serving as the residual connections~\cite{NGCF19}. \textcolor{black}{In Equation~(\ref{eq:semantic_agg}), we aggregate the semantics $\mathbf{s}^{(l)}_{h \leftarrow k}$ from $e_h$'s neighbors $\mathcal{N}_h$ to update its embedding $\mathbf{e}_h^{(l)}$ at layer $l$. As such, after iteratively propagating $L$ layers, $e_h$ is able to aggregate the semantics from its high-order (i.e., $L$-order) neighbors for refining its embedding, and the final entity embedding of $e_h$ is obtained by}:
\begin{equation}
\small
\mathbf{e}_h = \mathbf{e}^{(L)}_{h} \qquad \forall e_h \in \mathcal{G}_{(u,i)},
\end{equation}
\noindent where $\mathbf{e}^{(L)}_{h}$ is the output entity embedding from the last propagation layer $L$. As such, the layer-wise propagation mechanism sufficiently encodes the semantics and high-order topology of the subgraph into the well-learned entity embeddings. We further constitute an entity embedding matrix $\mathbf{H}_{(u,i)}$  for the subgraph $\mathcal{G}_{(u,i)}$ as follows:
\begin{equation}
\small
  \mathbf{H}_{(u,i)} = 
  \begin{bmatrix}
  \mathbf{e}_1, \mathbf{e}_2, \cdots, \mathbf{e}_n
  \end{bmatrix},
\end{equation}
\noindent where $n$ is the number of entities in $\mathcal{G}_{(u,i)}$. The entity embedding matrix $\mathbf{H}_{(u,i)} \in \mathbb{R}^{n\times d_e}$ is then fed into the subgraph embedding learning step to generate a holistic subgraph embedding for $\mathcal{G}_{(u,i)}$, so as to better characterize the complex user-item connectivity. 

\subsubsection{\textbf{Subgraph Embedding Learning}} 

\ 

\noindent Subgraph embedding learning generates the holistic subgraph embedding by attentively aggregating the entity embeddings, where the self-attention mechanism is employed to discriminate the importance of entities in the subgraph for enhanced embedding learning. 


\smallskip\noindent\textbf{Self-attention Mechanism. }For the subgraph $\mathcal{G}_{(u,i)}$ with $n$ entities, the self-attention mechanism takes the entity embedding matrix $\mathbf{H}_{(u,i)} \in \mathbb{R}^{n \times d_e}$ as input, and outputs the importance of entities in $\mathcal{G}_{(u,i)}$ via an attention network, formulated as:
\begin{equation}
\small
  \label{eq: singlehead}
  \mathbf{a}_{(u,i)} = softmax\left(\mathbf{w}_{2} \cdot \sigma (\mathbf{W}_{1}\mathbf{H}_{(u,i)}^T)\right),
\end{equation}
\noindent where $\mathbf{W}_{1} \in \mathbb{R}^{d_a \times d_e}$ and $\mathbf{w}_{2} \in \mathbb{R}^{d_a}$ are the trainable weight matrix and weight vector of the attention network, with $d_a$ as the hidden-layer size; $\sigma$ is the activation function of tanh; and $\mathbf{a}_{(u,i)} \in \mathbb{R}^{n}$ is the vector of attention scores for all entities in $\mathcal{G}_{(u,i)}$, normalized by the softmax function. As such, the $\mathbf{a}_{(u,i)}$ usually assigns higher importance to a specific set of entities, which is expect to reflect an aspect of the rich semantics within the subgraph. 
To this end, we generate multiple such attention score vectors that can focus on different aspects of $\mathcal{G}_{(u,i)}$ to capture its holistic semantics. This is implemented by extending the attention network in Equation (\ref{eq: singlehead}) into a multi-head attention, defined as: 
\begin{equation}
\small
  \label{eq: attmatrix}
  \mathbf{A}_{(u,i)} = softmax\left(\mathbf{W}_{2} \cdot \sigma (\mathbf{W}_{1}\mathbf{H}_{(u,i)}^T)\right),
\end{equation}

\noindent where $\mathbf{W}_{2} \in \mathbb{R}^{m \times d_a}$ is the trainable weight matrix, with $m$ as the number of attention heads; $\mathbf{A}_{(u,i)} \in \mathbb{R}^{m \times n}$ is the attention score matrix, where the row vectors discriminate the importance of entities in describing the different aspects of $\mathcal{G}_{(u,i)}$; and the number of attention heads $m$ thus controls how many aspects of $\mathcal{G}_{(u,i)}$ need to be concerned on for revealing the $u$'s preference towards $i$.

Based on the attention score matrix $\mathbf{A}_{(u,i)}$, we derive the subgraph embedding for $\mathcal{G}_{(u,i)}$ as follows:
\begin{equation}
\small
  \label{eq: subgraphemb}
  \mathbf{g}_{(u,i)} = f_g \left( \mathbf{A}_{(u,i)}\mathbf{H}_{(u,i)}\right).
\end{equation}
In Equation (\ref{eq: subgraphemb}), we multiply $\mathbf{A}_{(u,i)}$ and the entity embedding matrix $\mathbf{H}_{(u,i)}$ to compute $m$ weighted sums of entity embeddings $\mathbf{A}_{(u,i)}\mathbf{H}_{(u,i)}$, which are then aggregated by the function $f_g$ to obtain the holistic subgraph embedding $\mathbf{g}_{(u,i)}$. The aggregation function $f_g$ can be implemented with several operations (e.g., max-pooling, mean-pooling and vanilla attention mechanism), and we experimentally find out the mean-pooling operation outperforms the other choices, as will be introduced in Section \ref{sec: ablation}. It is worth pointing out the multiplication $\mathbf{A}_{(u,i)}\mathbf{H}_{(u,i)} \in \mathbb{R}^{m\times d_e}$ is size invariant with regard to the number of entities $n$ in the subgraph, and thus HAKG can learn embeddings for subgraphs with variable sizes. In sum, by leveraging a multi-head self-attention mechanism, the subgraph embedding $\mathbf{g}_{(u,i)}$ is able to provide a comprehensive description of the connectivity between $u$ and $i$, which is then utilized to predict the $u$'s preference towards $i$. 

\subsection{Preference Prediction}
Different from most prior works that merely utilize the user and item embeddings to predict user preferences, we content that the subgraphs connecting user-item pairs help reveal their underlying relationship, and thus are crucial for inferring the user preferences. Based on this assumption, we incorporate the user embedding $\mathbf{e}_u$, item embeddings $\mathbf{e}_i$, as well as the subgraph embedding $\mathbf{g}_{(u,i)}$ for better preference prediction, achieved by:
\begin{equation}
\label{eq:MLPscore}
    \widetilde{r}_{u,i} = MLP ([\mathbf{e}_u, \mathbf{g}_{(u,i)}, \mathbf{e}_i]),
\end{equation}
where $\widetilde{r}_{u,i}$ is the estimated ranking score for user $u$ on item $i$. In Equation~(\ref{eq:MLPscore}), we generate the prediction by feeding the concatenation of the three embeddings into a multi-layer perceptron, which has proven to be powerful in modeling the non-linear user-item connectivities~\cite{he2017neural}. Following the premise that neural networks can learn more abstractive features of data by using a small number of hidden units for higher layers~\cite{he2017neural}, we empirically implement a tower structure for the MLP, halving the layer size for each successive higher layer. We adopt ReLU as the activation function for hidden layers and sigmoid function for the output layer to control the estimated ranking score $\widetilde{r}_{u,i}$ into the range of $[0, 1]$.

\subsection{Model Optimization and Complexity} 

\smallskip\textbf{Objective Function.}
Following ~\cite{he2017neural}, we address the top-$N$ recommendation task as a binary classification problem, where the target value 1 means a user-item interaction is observed and 0 otherwise. Formally, we adopt the negative log-likelihood as the objective function, formulated by:
%
\begin{equation}
    \mathcal{J} = -\sum_{(u,i) \in \mathcal{R}^+} \log \widetilde{r}_{u,i} -\sum_{(u,j) \in \mathcal{R}^-}\log (1-\widetilde{r}_{u,j})
\end{equation}
%
where $\mathcal{R}^+$ and $\mathcal{R}^-$ denote the sets of observed and non-observed user-item interactions, respectively. 
For each observed interaction, we treat it as a positive instance, and pair it with one negative item that the user has not interacted with. We adopt mini-batch Adam~\cite{kingma2014adam} as the optimizer, and leverage the dropout strategy~\cite{NGCF19} to ease the over-fitting problem in optimizing deep neural network models. The optimization process is described by Algorithm \ref{alg:1}, \textcolor{black}{which is mainly composed of three parts: (1) Subgraph Construction (lines 1-3); (2) Hierarchical Attentive Subgraph Encoding (lines 6-9), where lines 6-7 describe the Entity Embedding Learning and lines 8-9 present the Subgraph Embedding Learning; and (3) Preference Prediction (line 10).}

\smallskip\noindent\textbf{Complexity Analysis.} 
The time cost of HAKG mainly comes from: (a) subgraph construction, (b) entity embedding learning, and (c) subgraph embedding learning. For (a), the computational complexity is $O(\vert\mathbf{R}\vert\overline{n})$, where $\vert\mathbf{R}\vert$ is the number of observed user-item interactions, and $\overline{n}$ is the average number of entities in the subgraphs. For (b), it takes $O(\vert\mathbf{R}\vert{L}{\overline{n} d_e^2)}$, where $L$ is the number of propagation layers, and $d_e$ is the embedding size of entities. For (c), the matrix multiplication has computational complexity $O(\vert\mathbf{R}\vert{m\overline{n}d_e)}$, where $m$ is the number of attention heads and usually less than $10$. The overall training complexity would be  $O(\vert\mathbf{R}\vert{L}{\overline{n} d_e^2)}$. In practice, $L, \overline{n}, d_e \ll \vert \mathbf{R}\vert$.
\textcolor{black}{Hence, the time complexity of HAKG is linear to the number of user-item interactions $\vert \mathbf{R}\vert$, and quadratic to the embedding size $d_e$.}

\begin{algorithm}[t]
\footnotesize
\caption{HAKG Optimization}\label{alg:1}
\KwIn{$\mathcal{G}, \mathbf{R}, K, L, d_e, d_t, d_r,d_a,m, \lambda,\gamma,$ max\_iter}
    \tcp {Subgraph Construction}
    \ForEach{($u, i$) pair in training set}{
        Sample $K$ paths $\mathcal{P}_{(u, i)}$ \;
        Assemble $\mathcal{P}_{(u, i)}$ into subgraph $\mathcal{G}_{(u,i)}$\;
      }
	\For{$iter=1;iter \le max\_iter;iter++ $}  
	{
		\ForEach{$(u, i)$ pair}{
        	\tcp {Entity Embedding Learning}
             Generate $\mathbf{e_h}$ based on Equations (1-4)\\
             Constitute $\mathbf{H}_{(u,i)}$ based on Equation (5)\\
            \tcp {Subgraph Embedding Learning}
            Generate $\mathbf{A}_{(u,i)}$  based on Equations (6-7)\\
            Generate $\mathbf{g}_{(u,i)}$ based on Equation (8)\\
            \tcp {Preference Prediction}
            Calculate $\tilde{r}_{u,i}$ based on Equation (9)\;
        }
		Update parameters by back propagation\;
	}
\end{algorithm}
\setlength{\textfloatsep}{7pt}

\section{Experiments and analysis}
We conduct extensive experiments on three real-world datasets with the goal of answering three research questions:

\smallskip\noindent\textbf{RQ1:} How does our proposed HAKG perform compared with state-of-the-art recommendation methods? Can it achieve better recommendation performance under different data sparsity levels?

\smallskip\noindent\textbf{RQ2:} How do the key designs of HAKG facilitate to improve the recommendation performance? 

\smallskip\noindent\textbf{RQ3:} How do different settings of hyper-parameters affect HAKG?

\subsection{Experimental Setup}\label{sec: setup}
\smallskip\textbf{Datasets.} 
Three benchmark datasets are utilized: (1) \textbf{MovieLens-1M}\footnote{https://grouplens.org/datasets/movielens/} is a widely used dataset in movie recommendations ~\cite{wang2018ripplenet}, which describes the user ratings towards movies ranging from
1-5; (2) \textbf{Last-FM}\footnote{https://grouplens.org/datasets/hetrec-2011/.} is \textcolor{black}{a music listening dataset collected from the Last.fm online music system}, where the tracks are viewed as items. We use the same version of this dataset as in ~\cite{wang2019kgat}; (3)
\textbf{Yelp}\footnote{http://www.yelp.com/dataset-challenge} records the user ratings on local business scaled from 1-5. Additionally, social relations among users and business attributes (e.g., category, city) are also included.

Note that as the main task of this paper is to perform the top-$N$ recommendation, we binarized the users' explicit ratings into implicit feedback for the three datasets. \textcolor{black}{Generally, there are two types of widely-used strategies to binarize the ratings into implicit feedback as ground truth, indicating whether the user has preference towards the item. The first type is only considering ratings no less than a threshold (e.g., 3) as implicit feedback~\cite{wang2018ripplenet, wang2019knowledge, wu2020diffnet++, liang2018variational}. The second type is to directly binarize the original user ratings into implicit feedback. We follow state-of-the-art methods~\cite{ wang2018explainable, jin2020efficient, he2017neural, hu2018leveraging} by using the second strategy in our study, which is the common setting in the area of recommender systems.} 
Besides user-item interactions, we merge more information into KGs for each dataset.
We combine MovieLens-1M with IMDb\footnote{https://www.imdb.com/.} as MI-1M by linking the titles and release dates of movies, so as to get side information about movies, such as genres, actors and directors. For Last-FM, we map
tracks into objects in the database called Freebase via title matching to get attributes of tracks, such as artists, engineers, producers, versions, types, contained\_songs, etc.
For Yelp, we extract knowledge from the social network and local business information network (e.g., category, city). Table \ref{tab:Dataset} summarizes the statistics of the three datasets.

\smallskip\noindent\textbf{Evaluation Protocols.}
We adopt \textit{leave-one-out}, which has been widely used in the previous efforts~\cite{hu2018leveraging,he2017neural}, to evaluate the recommendation performance. For each user, \textcolor{black}{we hold out her latest interaction as the test set, the second latest one for validation, and utilize the remaining data as training set}. Aligning with \cite{hu2018leveraging,he2017neural,wang2018explainable}, during testing phase, for each user, we randomly sample 100 items that the user has not interacted with as the negative items, and then rank her test item together with these negative items, so as to reduce the test complexity. \textcolor{black}{We adopt Hit@$N$, NDCG@$N$ and MRR@$N$ as evaluation metrics}, compute the three metrics for each test user and report the average score at $N = \{1,2,\cdots,15\}$ by following~\cite{he2017neural,wang2018explainable}. \textcolor{black}{Note that as the major task of this paper is to improve the recommendation accuracy, we only adopt the accuracy-based evaluation metrics and leave the evaluation based on diversity- and novelty-oriented metrics as future work~\cite{wu2019recent, kunaver2017diversity}.} The three metrics are defined as follows, and higher metric values generally indicate better ranking accuracy.
\begin{itemize}[leftmargin=*]
\item \textbf{Hit@$N$} measures whether the algorithm can recommend the test item within top-$N$ ranked list. The value would be 1 if the test item appears in the recommendation list, and 0 otherwise.
\item\textbf{NDCG@$N$} rewards each hit based on its position in the
ranked list, indicating how strongly an item is recommended. 
\item\textcolor{black}{\textbf{MRR@$N$} measures the reciprocal of the rank of each hit for each user.} 

\end{itemize}

\begin{table}[t]
\centering
  \caption{Statistics of the datasets.}\label{tab:Dataset}
  \vspace{0.06in}
  \footnotesize
  \renewcommand\arraystretch{0.95}
  \begin{tabular}{|c l| rrr|}
  \toprule
     &&\multicolumn{1}{c}{MI-1M}&\multicolumn{1}{c}{Last-FM}
     &\multicolumn{1}{c|}{Yelp}\\\midrule
     \multirow{4}{1.5cm}{\centering User-Item \\ Interactions} & \#Users 
     & 6,040    & 23,566    &37,940\\ \
     &\#Items 
     & 3,382    &48,123     &11,516\\ 
     &\#Interactions
     &756,684   &3,034,796  &229,178\\
     &Data Density
     &3.704\%   &0.268\% &0.052\%\\\midrule
     \multirow{3}{1.5cm}{\centering Knowledge\\Graphs}  
     &\#Entities 
     &18,920     &138,362    &46,606\\
     & \#Relation Types
     &10    &10          &6\\
     & \#Links 
     &968,038    &2,007,423    &302,937\\
     \bottomrule
\end{tabular}
\end{table}

\smallskip\noindent\textbf{Comparison Methods.}
We compare HAKG with four types of state-of-the-art recommendation methods:  
\begin{itemize}[leftmargin=*]
\item Plain CF-based with only user-item interactions: (a) \textbf{MostPop} - it recommends the most popular items to all users without personalization; (b) \textbf{BPRMF}~\cite{rendle2009bpr} 
- it is a classic algorithm that minimizes matrix factorization (MF) model with pairwise ranking loss for implicit feedback. (c) \textbf{NeuMF}~\cite{he2017neural}- it is a state-of-the-art deep
learning based method that combines MF with a MLP model for top-$N$ recommendation.
\item Direct-relation based method with KGs: (d) \textbf{CKE}~\cite{zhang2016collaborative} - it is a representative direct-relation based method, which leverages the embeddings derived from TransR~\cite{lin2015learning} to enhance MF. 
\item Path-based methods with KGs: (e) \textbf{FMG}~\cite{zhao2017meta} - it predefines various types of
meta-graphs and employs Matrix Factorization on the similarity matrix of each meta-graph
to make recommendations;
(f) \textbf{MCRec}~\cite{hu2018leveraging} - it is a state-of-the-art deep learning based method which extracts meta-paths between user-item pairs as their connectivities and encodes the paths with convolution neural networks; \textcolor{black}{(g) HAN~\cite{wang2019heterogeneous} - it is a recently proposed heterogeneous network embedding model, which leverages node-level and semantic-level attentions to learn the importance of entities and meta-paths\footnote{\textcolor{black}{We adapt HAN for the top-$N$ recommendation task. Specifically, we first learn entity embeddings by HAN and then employ the dot product of user and item embeddings to predict user preferences. For fair comparison, we optimize HAN with negative log-likelihood loss function.}};} (h) \textbf{RKGE}~\cite{sun2018recurrent} - it exploits the semantics of separate linear paths with the power of RNNs, and linearly aggregates the path embeddings for inferring user preferences; (i) \textbf{KPRN}~\cite{wang2018explainable} - it is a strong baseline which incorporates entity types into path embedding learning and performs weighted-pooling across the path embeddings to predict user preferences.
\item Propagation-based method with KGs: (j) \textbf{KGAT}~\cite{wang2019kgat} - it is a recently proposed state-of-the-art model that propagates information over the entire KG via GCNs to derive user and item embeddings, and use their inner products as estimated user preferences.
\end{itemize}

\smallskip\noindent\textbf{Parameter Settings. }The optimal parameter settings for all the
comparison methods are achieved by either empirical study or adopting the settings
as suggested by the original papers.
For HAKG, we apply a grid
search in $\{0.001, 0.002, 0.01, 0.02\}$ to find out the best learning rate $\gamma$; the optimal setting for $L_2$ regularization coefficient $\lambda$ is searched in $\{10^{-5}, 10^{-4}, 10^{-3}, 10^{-2}\}$. The best settings for other hyper-parameters are as follows: the batch size is 256; the embedding sizes of entity $d_e$, entity type $d_t$  and relation $d_r$ are set as $d_e=128, d_t=d_r=32$; the hidden layer size of the self-attention mechanism is set as $d_a = 128$, and the number of attention heads is set as $m = 5$ on the three datasets; and the size of predictive factors of MLP is set as 32. In addition, the number of sampled paths $K$ for each user-item pair is set to $15$ and the propagation layer is set as $L=2$ on the three datasets. We initialize the model parameters with the Xavier initializer~\cite{wang2019kgat}. The detailed analysis of hyper-parameter sensitivity will be introduced in Section 4.4.

\subsection{Performance Comparison (RQ1)}

In this subsection, we first compare the top-$N$ recommendation accuracy of HAKG with other state-of-the-art methods, and then
investigate how does the exploitation of KGs help alleviate the data sparsity issue. \textcolor{black}{We also compare the embedding learning ability and  the training time of HAKG with representative recommenders.}

\begin{figure*}[t]
\centering
\includegraphics[width=.98\textwidth]{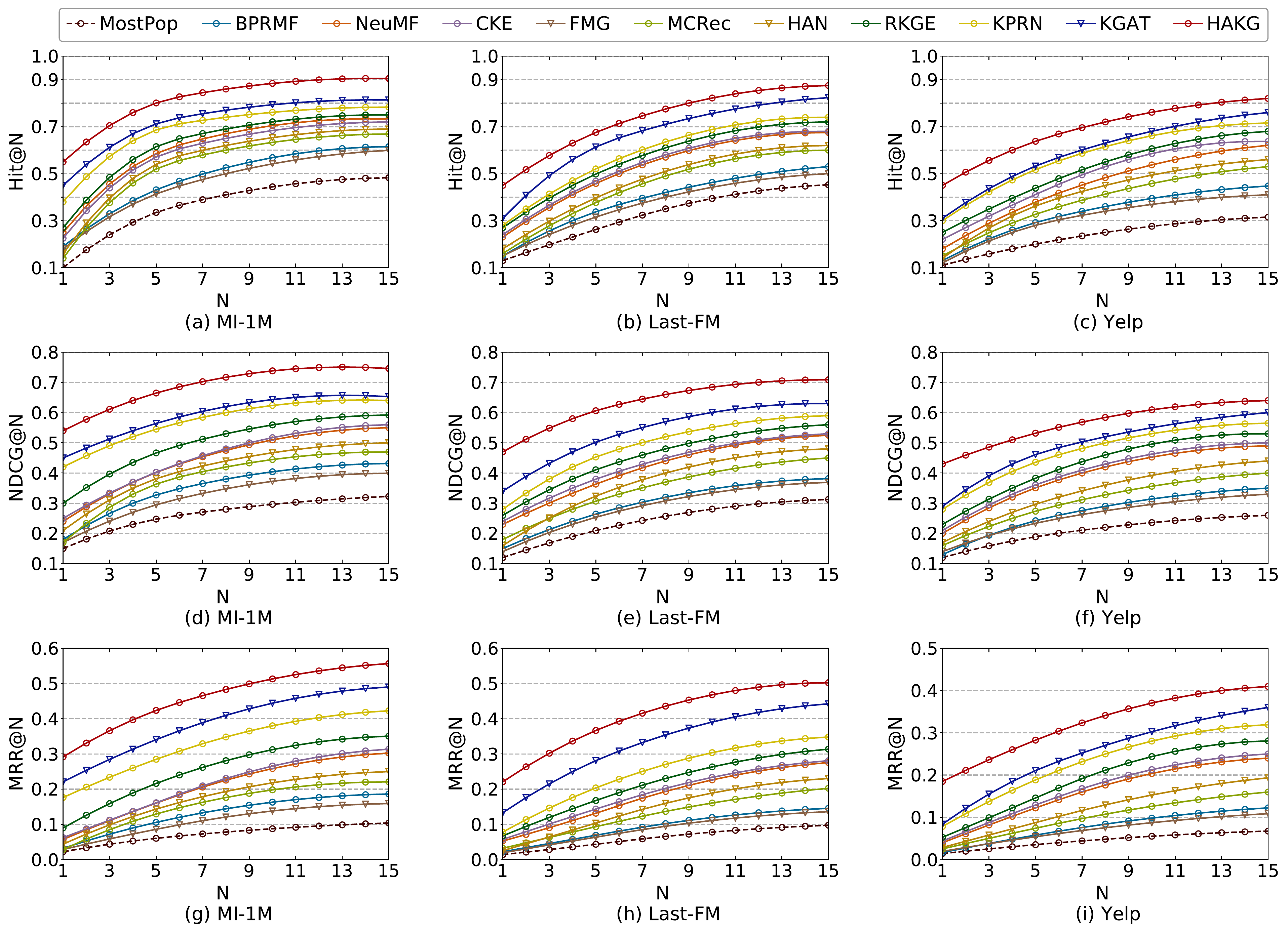}
\vspace{-0.12in}
\caption{Overall performance comparison on the three datasets w.r.t Hit@$N$, NDCG@$N$ and MRR@$N$ with $N=\{1,2,\cdots, 15\}$.
}\label{fig:allperformance}
\end{figure*}
\smallskip\noindent\textbf{Overall Performance Comparison.} 
Figure \ref{fig:allperformance} presents the overall performance comparison results on the three datasets w.r.t. Hit@$N$, NDCG@$N$ and MRR@$N$ with $N=\{1,2,\cdots,15\}$. We summarize the major findings as below.

First of all, among all the comparisons, most KG-aware methods outperform the plain CF-based methods on the three datasets, which indicates that incorporating KGs is able to greatly improve the recommendation performance. It is worth noting that NeuMF achieves better performance than BPRMF, implying the effectiveness of applying deep neural networks (i.e., MLP) to capture the non-linear user-item interactions. Moreover, the direct-relation based method CKE underperforms the path-based methods RKGE and KPRN, verifying that modeling only first-order relations fails to make full use of the information within KGs. It meanwhile indicates that path-based methods can better explore the user-item connectivities by explicitly encoding the paths to infer user preferences. 

Second, in terms of path-based baselines, both KPRN and RKGE are based on the automatically mined semantic paths, while FMG and MCRec heavily rely on the quality of handcrafted meta-paths. The performance of KPRN and RKGE far exceeds that of FMG, MCRec and HAN, verifying that the pre-defined features fail to uncover all potential relations between entities. \textcolor{black}{In addition, HAN achieves unsatisfactory performance on the three datasets. This might because HAN forgoes all intermediate entities along the meta-path, which leads to information loss and is insufficient to capture the holistic semantics of user-item connectivities. The observation is consistent with ~\cite{hu2018leveraging}. } Furthermore, KPRN performs better than RKGE as it additionally takes the different importance of paths into consideration via a weighted-pooling operation. In addition, the most recent propagation-based method KGAT outperforms the path-based methods KPRN and RKGE, which implies the superiority of 
graph neural networks in modeling the high-order connectivities via information propagation. However, KGAT has a large gap with HAKG on the three datasets, which verifies that propagating information over the full KG is less efficient
to model the pair-wise user-item connectivity for accurate recommendation, as it would introduce noise that is irrelevant to the specific connectivity into embedding learning process. 

Overall, our proposed HAKG consistently achieves the best performance among all the comparisons on the three datasets w.r.t. all evaluation metrics.
The relative improvements achieved by HAKG over the runner up w.r.t. Hit, NDCG amd MRR are 10.7\%, 10.1\%, 13.6\% on average, respectively. The improvements are statistically significant proven by commonly used paired t-test over 20-round results~\cite{NGCF19} with $p$-value$<0.05$. We attribute the superiority of HAKG in recommendation performance to two core advantages: (1) the expressiveness of subgraphs in characterizing user-item connectivities; and (2) the effectiveness of hierarchical attentive embedding learning procedure in encoding the subgraphs for revealing user preferences.

\begin{figure*}[t]
\centering
\includegraphics[width =0.99\textwidth]{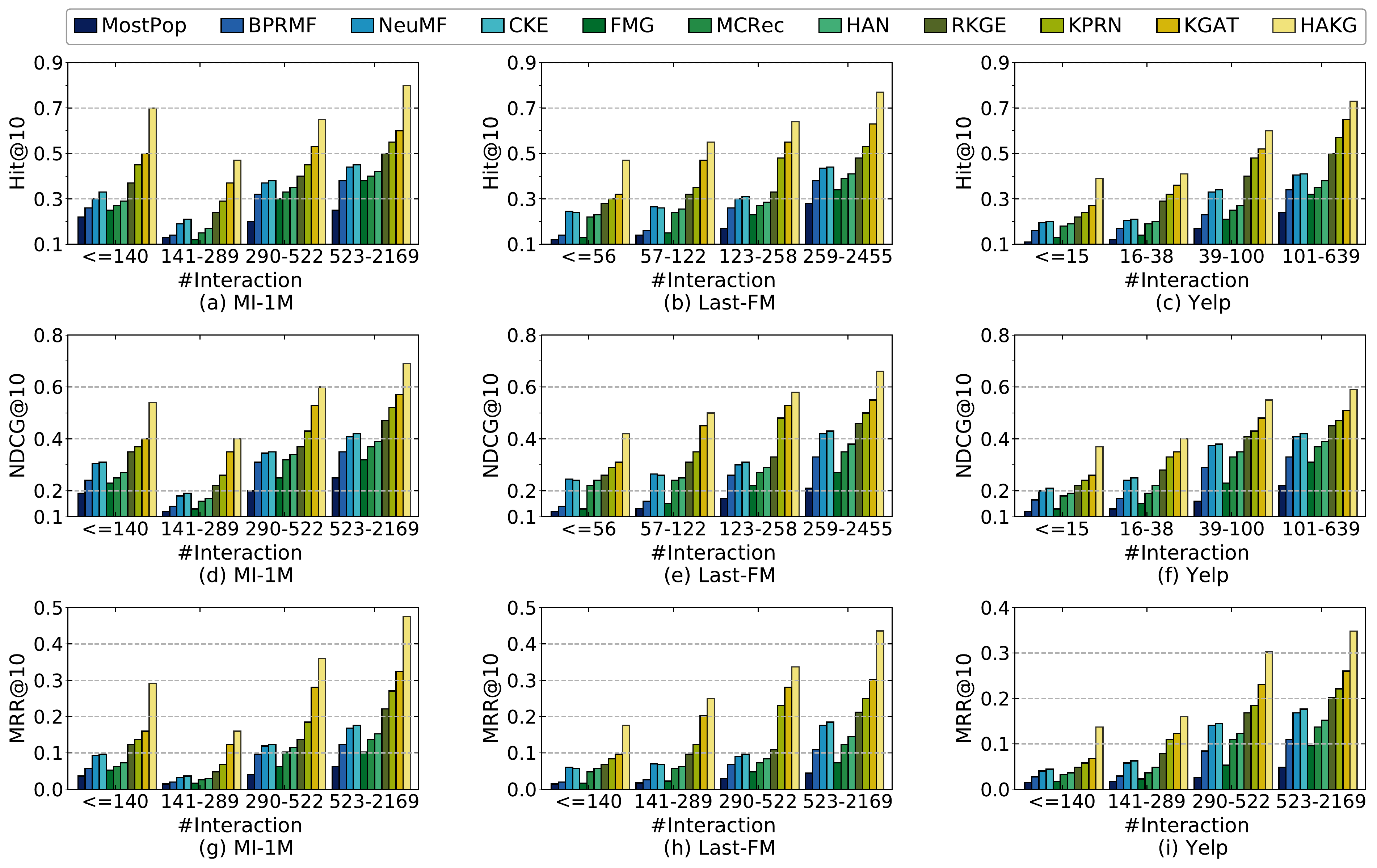}
\vspace{-0.12in}
\caption{Performance comparison on user groups with different sparsity levels on the three datasets w.r.t. Hit@10, NDCG@10 and MRR@10. Similar trends can be observed on the other settings of $N$.} 
\label{fig:coldstart}
\end{figure*}
\smallskip\noindent\textbf{Performance Comparison w.r.t. Data Sparsity Levels. }We further investigate whether exploiting
the rich information within KGs helps alleviate the data sparsity issue.
Towards this end, we perform experiments over user groups of different sparsity levels. \textcolor{black}{In particular, following state-of-the-art methods~\cite{wang2019kgat, NGCF19}, we divide the test users into four groups based on interaction number per user in the training set, and each group has the same total interactions. The density of each user group then can be calculated by $\frac{\# \text{total interactions}}{\# \text{users} \times \# \text{items}}$.
Specifically, the density of the four groups are 1.5\%, 5.1\%, 9.6\%, 18.9\% in MI-1M; 0.05\%, 0.17\%, 0.35\%, 0.91\% in Last-FM; and 0.02\%, 0.16\%, 0.43\%, 1.19\% in Yelp. That is, the sparsity level of the four groups decrease gradually,
where the first group is the sparsest one.} Figure~\ref{fig:coldstart} illustrates
the results w.r.t. Hit@10, NDCG@10 and MRR@10 on different user groups in the three datasets, and similar trends can be observed on the other settings of $N$. 

We can find that: first of all, HAKG consistently outperforms all comparisons w.r.t. different user groups on the three datasets, and meanwhile the improvements become more significant as the interactions of user groups are sparser. This verifies the advantage of modeling user-item connectivities via the expressive subgraphs, which enriches the embeddings of inactive users by exploiting the information within KGs. In addition, KGAT achieves unsatisfactory performance on the densest user group (i.e., the
fourth group) in MI-1M and Last-FM, suggesting that propagating information over the entire KG cannot distill the salient relations for accurate recommendation. 
\textcolor{black}{Contrarily, the subgraph shows its superiority in retaining only the relevant entities and relations for modeling the specific user-item connectiv·ity}
. 

\begin{figure*}[t]
\centering
\includegraphics[width=0.95\textwidth]{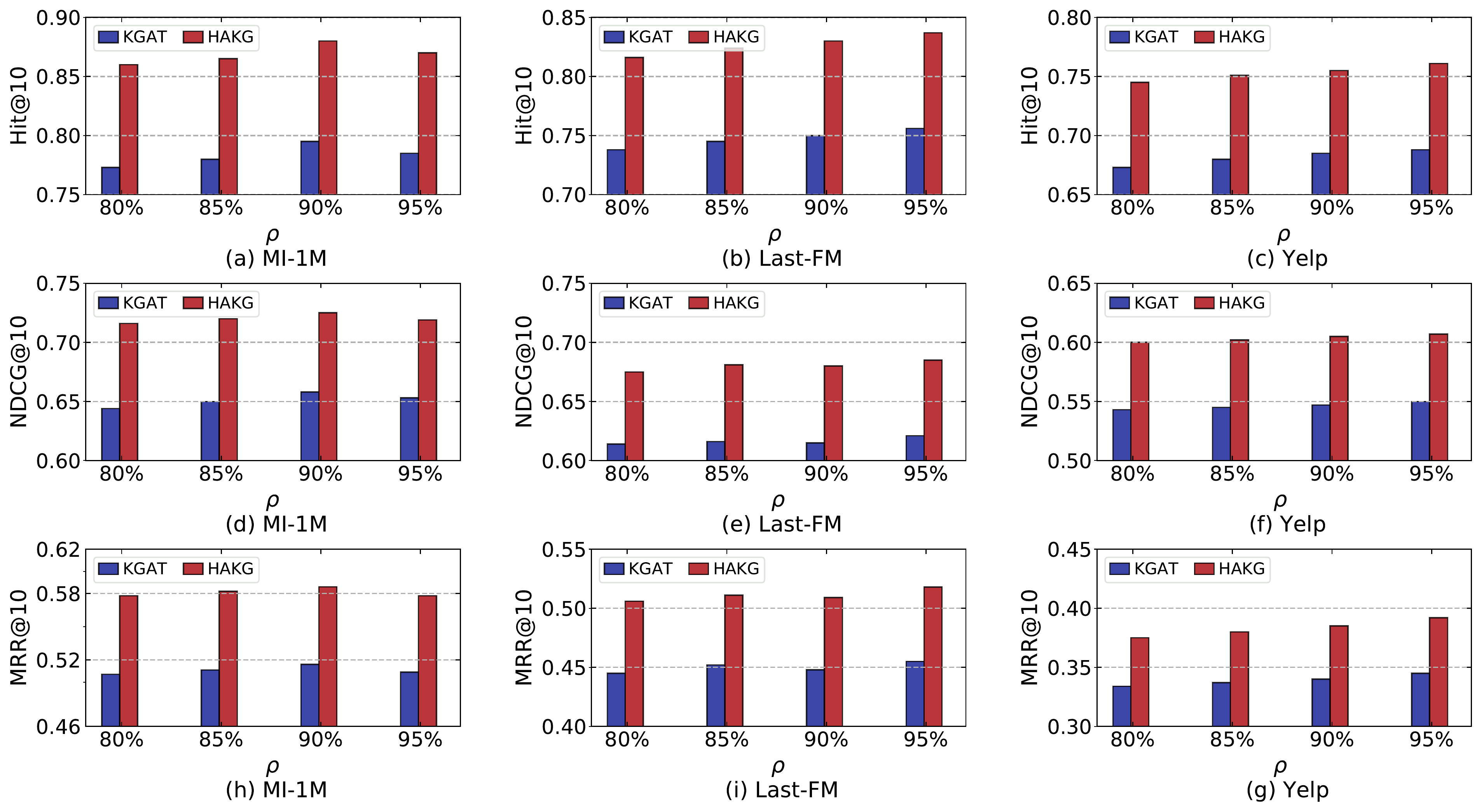}
\vspace{-0.08in}
\caption{Performance comparison between our HAKG and the strongest baseline KGAT on different proportions of training set $\rho$ w.r.t Hit@10, NDCG@10 and MRR@10 on the three datasets.}
\label{fig: test_portion}
\end{figure*}

\smallskip\noindent\textbf{Performance Comparison on Different Data Splitting Methods. }\textcolor{black}{We investigate whether different data splitting methods would impact the recommendation performance of HAKG. To this end, we further conduct experiments by adopting the split-by-ratio strategy. In particular, we randomly select a portion (i.e. $\rho = \{80\%, 85\%, 90\%, 95\%\}$) of historical interactions of each user to constitute the training set, and the rest interactions are treated as the test set (i.e. $1-\rho = \{5\%, 10\%, 15\%, 20\%\}$), correspondingly. From the training set, we randomly select 10\% of interactions as validation set to tune hyper-parameters. Figure~\ref{fig: test_portion}  presents  the  performance comparison results between HAKG and the strongest baseline KGAT w.r.t. Hit@10, NDCG@10 and MRR@10 on the three datasets, and similar trends can be observed on the other settings of $N$. We can find that HAKG consistently achieves the best performance across all evaluation metrics w.r.t. different settings of $\rho$. The relative improvements achieved by HAKG over the runner up w.r.t. different settings of $\rho$ are 10.7\%, 10.5\%, 10.9\%, 10.8\% (Hit@10); 10.2\%, 10.4\%, 10.6\%, 10.5\% (NDCG@10); and 13.6\%. 13.4\%, 13.2\%, 13.3\% (MRR@10), on the three datasets averagely. This firmly verifies the effectiveness of HAKG in generating better recommendation regardless of the data splitting methods.}

\begin{figure*}[t]
\centering
\includegraphics[width=1\textwidth]{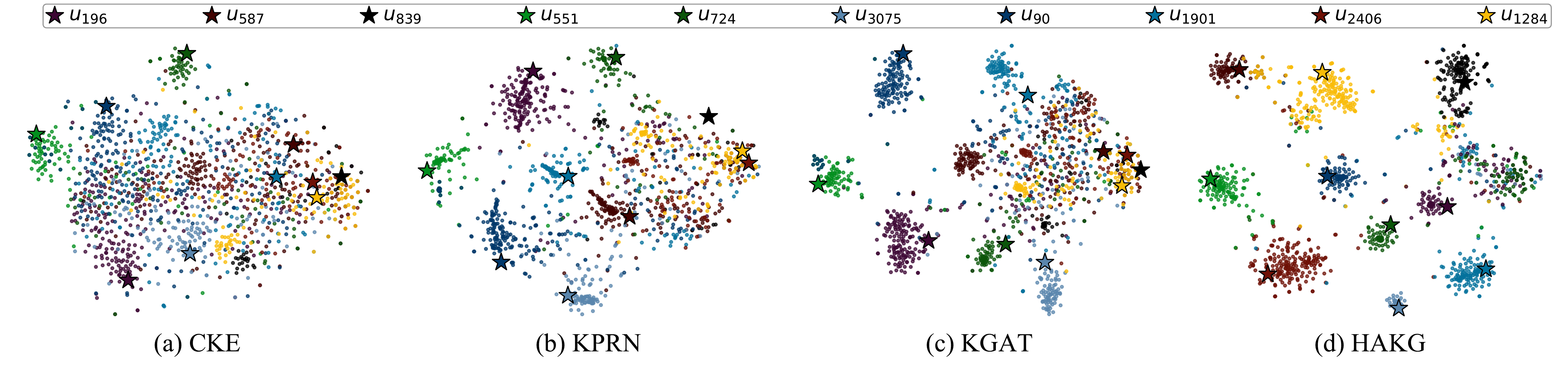}
\vspace{-0.4in}
\caption{Visualization of the t-SNE~\cite{maaten2008visualizing} transformed
embeddings. Each `$\star$'
represents a user from MI-1M dataset and the dots
with the same color denote her interacted items. Numbers in the legend are user IDs.
}\label{fig:visual}
\end{figure*}

\smallskip\noindent\textbf{\textcolor{black}{Embedding Learning Ability. }}\textcolor{black}{We examine whether HAKG is able to learn better embeddings for inferring user preferences towards items. To achieve this, following state-of-the-arts~\cite{NGCF19}, we randomly select ten users and their interacted items from MI-1M. We then visualize their embeddings learned via representative recommenders, namely CKE, KPRN, KGAT and HAKG, as shown in Figure~\ref{fig:visual}. Comparing Figure~\ref{fig:visual}(a-c) with Figure~\ref{fig:visual}(d), we can observe that the proposed HAKG, which exploits the user-item connectivities via expressive subgraphs, advances the other three baselines in learning more accurate embeddings: each user and her interacted items are close to each other in the embedding space, and meanwhile the items of the same user tend to form a cluster. }

\begin{table}[t]
\footnotesize
\centering
\caption{Comparisons of training time (second/hour [s/h]).
`Average' is the average training time for each epoch, `Total' denotes the total training time to converge.}  
\label{tab:runnningtime}
\vspace{0.05in}
 \resizebox{0.6\textwidth}{!}{
 \begin{tabular}{|c|c c |cc| cc|}
  \toprule
      \multirow{2}{*}{}&\multicolumn{2}{c|}{\textbf{MI-1M}}&\multicolumn{2}{c|}{\textbf{Last-FM}}&\multicolumn{2}{c|}{\textbf{Yelp}}\\\cmidrule{2-7}
     & Average & Total& Average & Total& Average & Total \\\midrule
    BPRMF
    &61.7s&1.5h
    &72.7s&2.0h
    &9.5s&0.2h\\
    NeuMF
    &67.5s&1.8h
    &189.5s&5.3h
    &12.3s&0.3h\\
     CKE
    &70.1s&1.9h
    &129.1s&3.6h
    &13.7s&0.4h\\
     KPRN
     &1.9h&21.4h
     &4.1h&45.7h
     &0.8h& 8.2h\\
     KGAT
    &106.8s&3.1h
    &233.7s&6.5h
    &20.6s&0.6h \\
    HAKG
    &216.8s&6.2h
    &364.1s&9.8h
    &62.5&1.7h \\
    \bottomrule
     \end{tabular}}
\end{table}

\smallskip\noindent\textbf{Training Time Comparison. }We compare the training time of HAKG with representative state-of-the-art recommenders, including the plain CF-based methods (BPRMF, NeuMF) and KG-aware methods (CKE, KPRN, KGAT). All the experiments are conducted on a single GPU of Nvidia GeForce GTX 1080 Ti. From the experimental results in Table \ref{tab:runnningtime}, we can observe that HAKG yeilds comparable complexity to the propagation-based method KGAT, and \textcolor{black}{it is much efficient than the path-based method KPRN with an relative improvement of 76.3\% on average for the three datasets}.

\subsection{Ablation Analysis of HAKG (RQ2)}
\label{sec: ablation}
In this subsection, we
conduct an ablation analysis to investigate the efficacy of the key designs of HAKG, including (1) entity type and relation; (2) self-attention mechanism; (3) aggregation function $f_g$; and (4) subgraph embedding.

\smallskip\noindent\textbf{Impact of Entity Type and Relation.}
HAKG incorporates both the entity types and relations into entity embedding learning process. To study their respective impact, we consider two variants of HAKG, including HAKG-t that removes the entity type embedding during embedding initialization in Equation (\ref{equ:typeembedding}), and HAKG-r that removes the relation embedding during semantics propagation in Equation (\ref{equ: propgation}). 
From the results in Table \ref{tab:variants}, we notice that both the two variants underperform HAKG, which implies that the incorporation of heterogeneous entity types and relations facilities to improve the recommendation performance, by providing the rich semantic information for learning enhanced entity embeddings. 

\smallskip\noindent\textbf{Impact of Self-attention Mechanism.} 
HAKG employs the self-attention mechanism to learn entity importance during the subgraph embedding learning process. To investigate its impact, we compare HAKG with the variant HAKG-a, which removes the self-attention mechanism in Equation~(\ref{eq: attmatrix}) and combines entity embeddings with the mean-pooling operation. As shown in Table~\ref{tab:variants}, the performance of HAKG-a evidently declines on all three datasets, with a decrease of 4.85\% on average. This \textcolor{black}{verifies that the self-attention mechanism can assist in better user preference inference by specifying the varying importance of entities in the subgraph.}

\smallskip\noindent\textbf{Impact of Aggregation Function.} 
HAKG deploys the function $f_g$ to aggregate the weighted sums of entity embeddings $\mathbf{A}_{(u,i)}\mathbf{H}_{(u,i)}$ into the subgraph embedding in Equation (\ref{eq: subgraphemb}). In HAKG, $f_g$ is implemented with the mean-pooling operation, and here we evaluate the performance of other operations by considering two variants HAKG$_{max}$ and HAKG$_{att}$, which implement $f_g$ with max-pooling operation and vanilla attention mechanism, respectively. From the results in Table \ref{tab:variants}, we can observe that HAKG$_{max}$ performs the worst on all three datasets. This is mainly because the max-pooling operation only preserves the most important features, and thus leading to information loss. Furthermore, HAKG yields better performance than HAKG$_{att}$. This implies that an average combination is sufficient for learning effective subgraph embeddings, which can be attributed to the well learned importance of entities from the self-attention mechanism.

\begin{table*}[t]
  \centering
  \footnotesize
  \caption{Ablation analysis on the three datasets w.r.t. Hit@10 and NDCG@10. `Decrease' denotes the relative performance decrease of corresponding variants compared with HAKG. 
 }  \label{tab:variants}
 \vspace{0.05in}
  \renewcommand\arraystretch{1}
  \resizebox{\textwidth}{!}{
  \begin{tabular}{|c|cc cc |cccc| cccc|}
  \toprule
     &\multicolumn{4}{c|}{\textbf{MI-1M}}&\multicolumn{4}{c|}{\textbf{Last-FM}}&\multicolumn{4}{c|}{\textbf{Yelp}}\\\cmidrule{2-13}
     & Hit&Decrease& NDCG&Decrease&Hit&Decrease& NDCG&Decrease& Hit&Decrease& NDCG&Decrease \\\midrule
     \textbf{HAKG}
     &\textbf{0.883}&--&\textbf{0.725}& --
     &\textbf{0.837}&--&\textbf{0.685}& -- 
     &\textbf{0.761}&--&\textbf{0.609}&--\\
     HAKG-t
     &0.869&-1.59\%&0.713&-1.65\%
     &0.821&-1.91\%&0.672&-1.89\%
     &0.750&-1.45\%&0.601&-1.31\%\\
     HAKG-r
     &0.863&-2.27\%&0.708&-2.34\%
     &0.819&-2.15\%&0.668&-2.48\%
     &0.745&-2.10\%&0.596&-2.13\%\\
     HAKG-a 
     &0.837&-5.21\%&0.686&-5.38\%
     &0.798&-4.66\%&0.653&-4.67\%
     &0.726&-4.60\%&0.581&-4.60\%\\
     HAKG-g
     &0.812&-8.04\%&0.671&-7.45\%
     &0.774&-7.53\%&0.643&-6.13\%
     &0.690&-9.33\%&0.561&-7.88\%\\
     \midrule
     HAKG$_{max}$
     &0.850&-3.73\%&0.698&-3.72\%
     &0.813&-2.87\%&0.665&-2.92\%
     &0.735&-3.42\%&0.587&-3.61\%\\
     HAKG$_{att}$
     &0.853&-3.40\%&0.701&-3.31\%
     &0.816&-2.51\%&0.667&-2.63\%
     &0.739&-2.89\%&0.592&-2.79\%\\
    \bottomrule
     \end{tabular}
}
\vspace{-0.02in}
\end{table*}

\smallskip\noindent\textbf{Impact of Subgraph Embedding.} HAKG fuses the learned subgraph embedding into user preference prediction in Equation (\ref{eq:MLPscore}). To verify its effectiveness, we compare HAKG with the variant HAKG-g by removing the subgraph embedding, that is, HAKG-g performs prediction with only user and item embeddings. As observed in Table \ref{tab:variants}, HAKG-g consistently performs the worst in all metrics across the three datasets. This confirms that the incorporation of subgraph embeddings into user preference prediction can boost the recommendation performance. \textcolor{black}{We also examine the effects of our subgraph construction strategy by generating subgraphs via BFS and DFS. For the smallest Yelp dataset, both BFS and DFS take about 15 days to construct subgraphs for all user-item pairs; while our subgraph construction strategy costs around 21 minutes, 52 minutes and 324s on the three datasets, respectively. This demonstrates that the traditional graph mining methods are computationally prohibitive by enumerating all the paths between a user-item pair; whereas our subgraph construction strategy can scale to large-scale real-world datasets via efficient path sampling and assembling process. }

\subsection{Parameter Sensitivity (RQ3)}
\label{parasensitivity}
We finally study how the representative hyper-parameters affect the performance of HAKG, including the number of sampled paths $K$, the propagation layers $L$, the multi-heads of self-attention mechanism $m$, the embedding sizes of entity $d_e$ and entity type $d_t$. In the following experiments, we vary the hyper-parameter being test while using the optimal settings for the rest parameters as introduced in Section \ref{sec: setup}. 

\begin{figure}[t]
\vspace{-0.015in}
\centering
\includegraphics[width =0.7\textwidth]{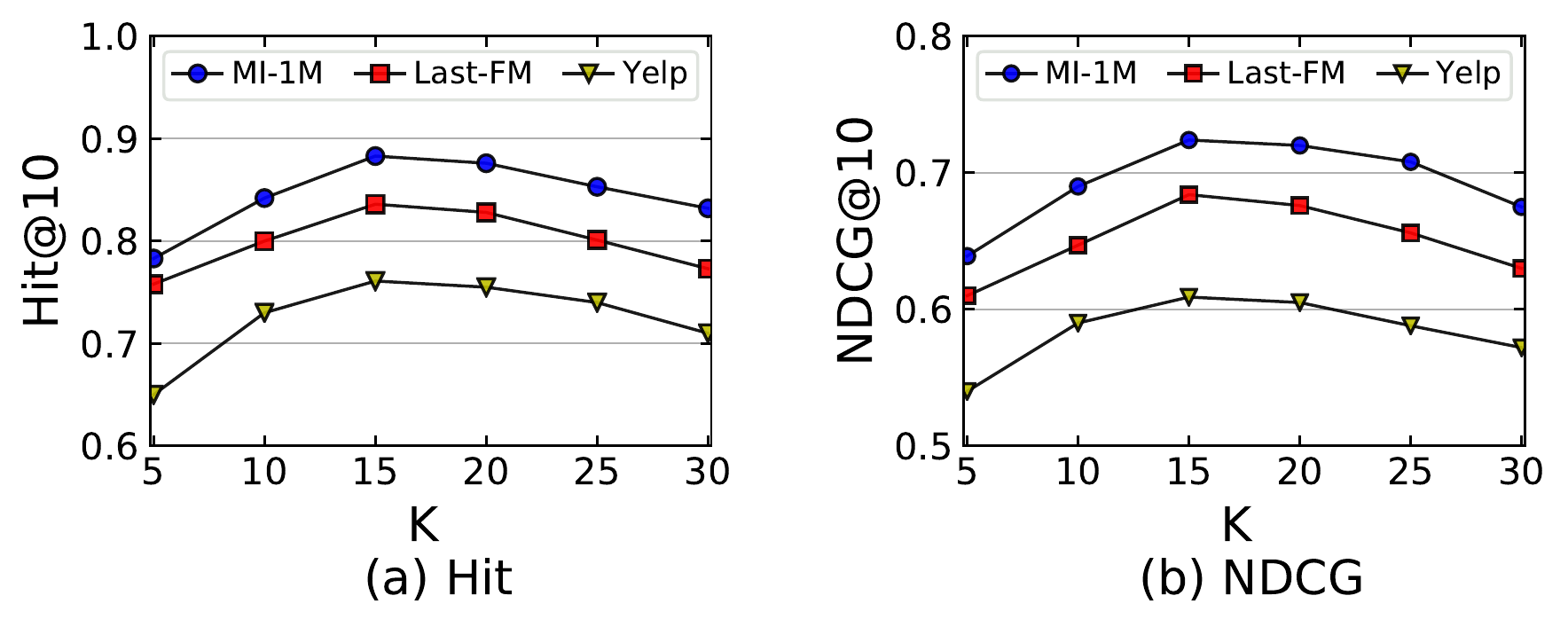}
\vspace{-0.2in}
\caption{Parameter sensitivity w.r.t. the number of sampled paths $K$ on Hit@10 and NDCG@10.}
\label{fig: path_num}
\end{figure}
\begin{figure}[t]
\vspace{-0.015in}
\centering
\includegraphics[width =0.69\textwidth]{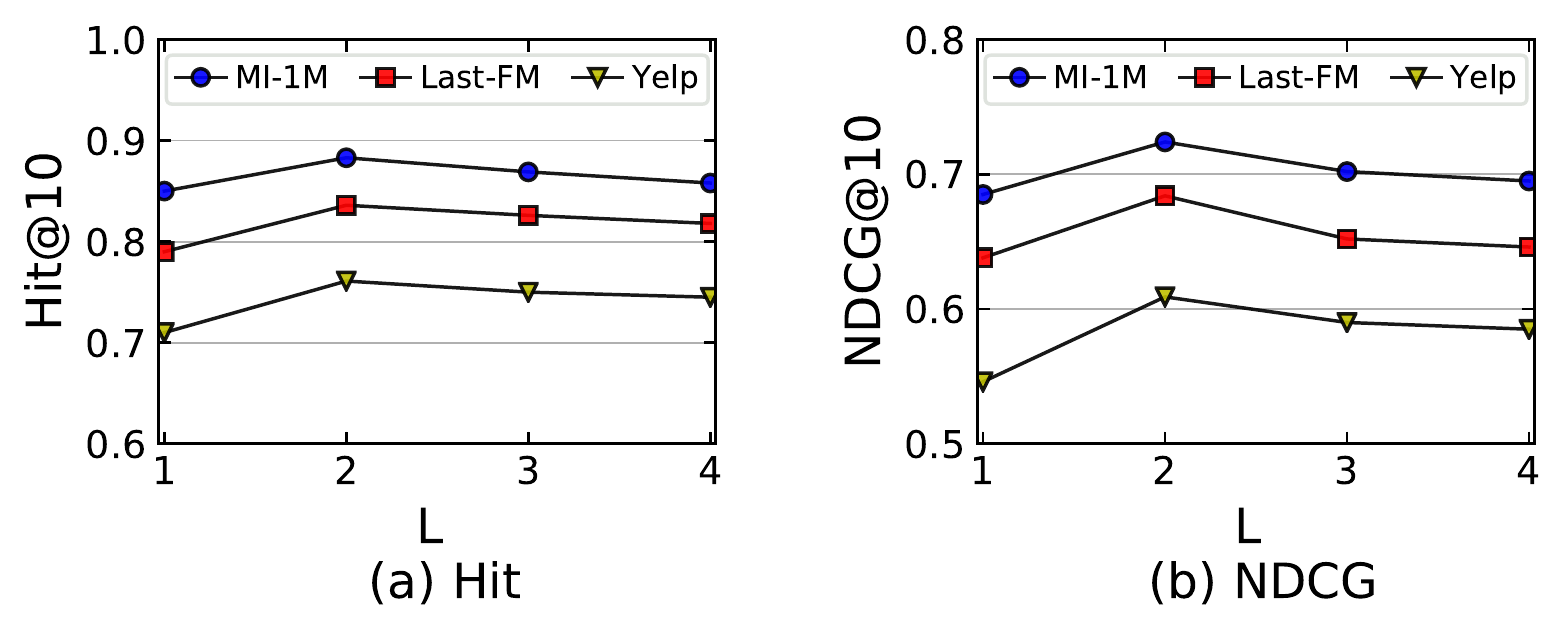}
\vspace{-0.25in}
\caption{Parameter sensitivity w.r.t. the number of propagation layers $L$ on Hit@10 and NDCG@10.}
\label{fig:L}
\end{figure}

\smallskip\noindent\textbf{Number of Sampled Paths. }We vary the number of sampled path $K$ in the range of $\{5, 10, 15, 20, 25, 30\}$ to study its impact on recommendation accuracy.
As shown in Figure \ref{fig: path_num}, we observe that as $K$ increases, the performance of HAKG improves at first since more paths could help encode the rich knowledge from KGs. The optimal performance is obtained with $K=15$ on the three datasets, while gradually drops with further increase of $K$. This implies that too much integration of paths would introduce noise even degrade the recommendation performance of HAKG.

\smallskip\noindent\textbf{Propagation Layers.} We study the influence of propagation layers $L$ by varying $L$ in the range of $\{1, 2, 3, 4\}$. The results are presented in Figure \ref{fig:L}, where we observe that HAKG achieves a better performance with $L=2$ over $L=1$. 
This suggests that increasing the depth of propagation contributes to an effective modeling of the high-order connectivities. 
However, the performance decreases when stacking more layers (i.e., $L=3$ and $L=4$). This implies that considering second-order relations among entities could be sufficient to help capture the semantics of subgraphs, while stacking more propagation layers may lead to over-fitting~\cite{wang2019knowledge}.

\begin{figure}[t]
\centering
\includegraphics[width =0.7\textwidth]{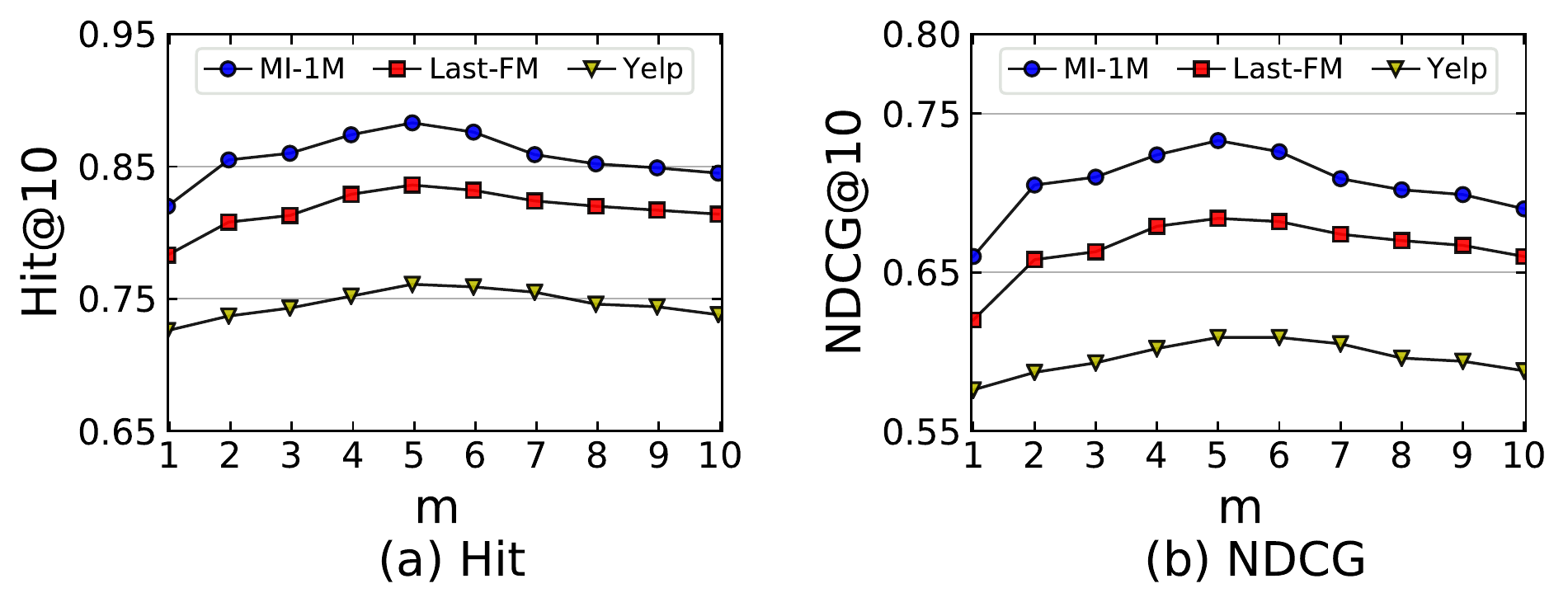}
\vspace{-0.25in}
\caption{Parameter sensitivity w.r.t. multi-heads of self-attention mechanism $m$ on Hit@10 and NDCG@10.}
\label{fig: m}
\end{figure}
\begin{figure}[t]
\centering
\includegraphics[width =.7\textwidth]{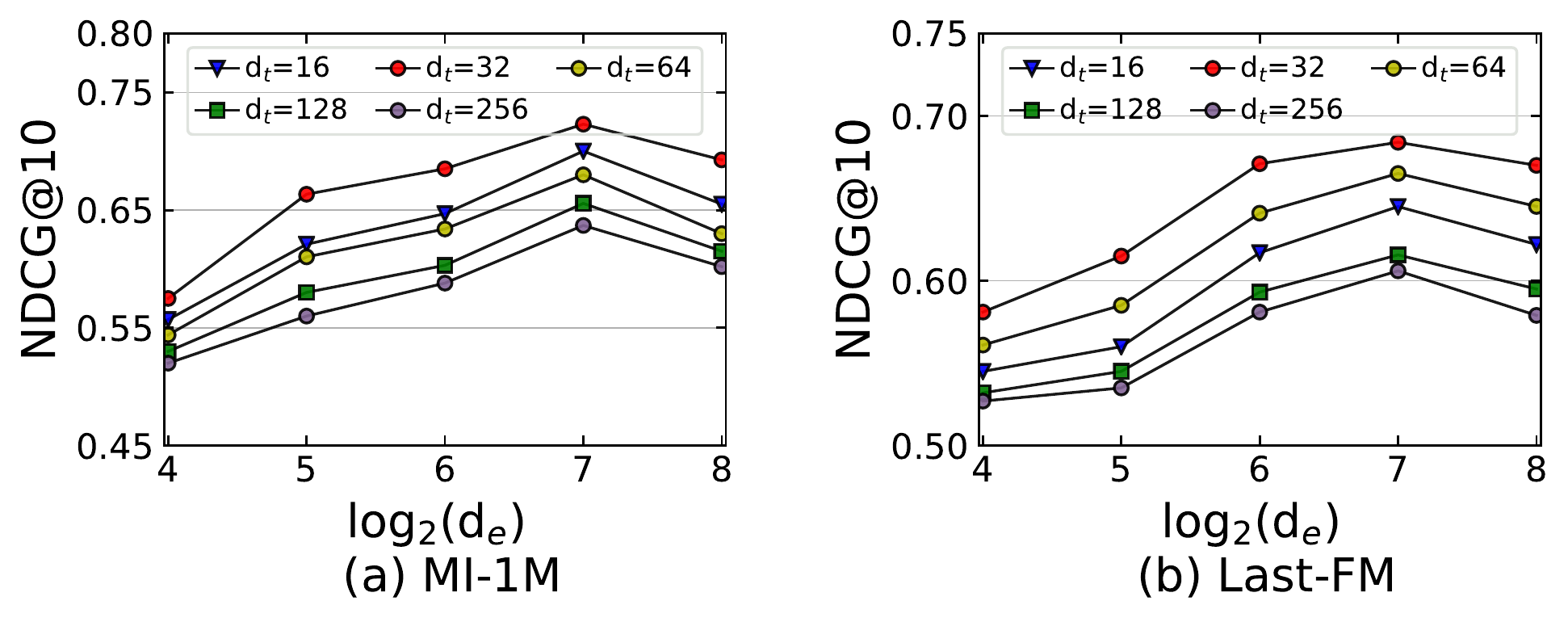}
\vspace{-0.25in}
\caption{Parameter sensitivity w.r.t. the embedding sizes of entity $d_e$ and entity type $d_t$ on NDCG@10 for MI-1M and Last-FM datasets. \textcolor{black}{We use $log_2(d_e)$ as x-axis to ensure the linear increase of the embedding size.} }
\label{fig: d_d'}
\end{figure}

\smallskip\noindent\textbf{Multi-heads of Self-attention Mechanism.}
The ablation analysis in Section 4.3 has verified the self-attention mechanism is beneficial for improving the recommendation performance. Here, we mainly evaluate how the number of attention heads $m$ impacts the performance of HAKG by varying $m$ in $\{1,2,\cdots,10\}$. From the results in Figure \ref{fig: m}, we can find that the performance of HAKG first climbs up and yeilds the best with $m=5$, while declines with the further increase of $m$.  This suggests that the setting of $m$ indeed effects the recommendation accuracy, by controlling how many different aspects of the subgraph need to be focused on for revealing user preferences. In particular, a too small number of $m$ is insufficient to capture the holistic semantics from the multiple aspects of the subgraph, while using too many aspects would introduce the redundancy information, thus limiting the performance.

\smallskip\noindent\textbf{Embedding Sizes.} We examine how the sizes of entity embedding $d_e$ and entity type embedding $d_t$ affect HAKG, by testing all the combinations of $d_e$ and $d_t$ in the range of $\{16, 32, 64, 128, 256\}$. Due to space limitation, we only report the performance w.r.t NDCG@10 on MI-1M and Last-FM datasets, shown by Figure~\ref{fig: d_d'}. Similar trends can be observed on Yelp dataset and other metrics. From the results, we note that, given a fixed size of entity type embeddings (e.g., $d_t=16$), the performance first improves with the increase of entity embedding size $d_e$, and the best performance is achieved when $d_e=128$. This suggests that embeddings with larger size can remarkably help encode useful information for effective embedding learning. The performance however, drops a lot with further increase of $d_e$ (i.e., $d_e=256$), as oversized embeddings may over-represent the entities, thus introducing noise. Similar trends are also possessed by $d_t$, and the optimal settings of the embedding sizes are $d_e=128$ and $d_t=32$ on the three datasets.

\section{CONCLUSIONS AND FUTURE WORK}
We propose a novel hierarchical attentive knowledge graph embedding framework to exploit the user-item connectivities in KGs for enhanced recommendation. We harvest the heterogeneous subgraphs between the user-item pairs for comprehensively characterizing their connectivities, and design a hierarchical attentive embedding learning procedure to effectively encode the subgraphs for revealing user preferences. Specifically, the
layer-wise propagation mechanism sufficiently encodes both the semantics and topology of subgraphs into the learned entity embeddings, and the self-attention mechanism helps discriminate the importance of entities for effective subgraph embedding learning.
Extensive experiments over three real-world datasets demonstrate the superiority of HAKG against the state-of-the-art recommendation methods, as well as its potential in easing the data sparsity issue.

\textcolor{black}{In the future, we plan to extend our work in two directions. The first direction is to further involve the temporal context into KGs.  In real world, user preferences usually evolve over time, which can be affected by dynamic user inclinations, item perception and popularity. Temporal context hence has been recognized as a crucial type of information for modeling the dynamic user preferences. Thus, we would like to 
adapt the proposed HAKG for dynamic recommendation by incorporating the temporal context into KGs. Specifically, we can model the user clicked items in a sequence 
as a directed graph, where each link $(i_k,i_j)$ represents a user visits items $i_k$ and $i_j$ consecutively. The 
sequential interactions
then can be seamlessly
integrated with the KG as a unified graph, which accommodates both the transition relationship between items and the item side information for better understanding the user preferences. Another promising direction is to explore the KGs for diversified recommendation, which could greatly facilitate user satisfaction~\cite{wu2019recent}. 
Our proposed HAKG has revealed improved 
accuracy by exploiting KGs in recommendation.
This motivates us to further incorporate the KGs for enhancing the 
diversity of recommendations.
In particular, we will attempt to combine the KG embeddings with determinantal point processes~\cite{wu2019recent}, so as to better balance recommendation accuracy and diversity.}


\bibliography{main}

\end{document}